\pdfoutput=1
\documentclass[aps,prl,twocolumn,groupedaddress,showpacs]{revtex4-1}

\usepackage{amssymb,amsmath}
\usepackage{graphics}
\usepackage{dsfont}

\def\kr{k_R}                            				
\def\Er{E_R}                            				
\def\Rb87{^{87}\mathrm{Rb}}                     
\def\K40{^{40}\mathrm{K}}                    		

\def\ket#1{\mathinner{|{#1}\rangle}}

{\catcode`\|=\active
  \gdef\Braket#1{\left<\mathcode`\|"8000\let|\BraVert {#1}\right>}}
\def\BraVert{\egroup\,\mid@vertical\,\bgroup}

\begin{document}

\title{A Raman-induced Feshbach resonance in an effectively single-component Fermi gas}


\author{R. A. Williams, M. C. Beeler, L. J. LeBlanc, K. Jim{\'e}nez-Garc{\'i}a}
\author{I. B. Spielman}
\affiliation{Joint Quantum Institute, National Institute of Standards and Technology, and University of Maryland, Gaithersburg, Maryland, 20899, USA}

\date{\today}

\begin{abstract}
Ultracold gases of interacting spin-orbit coupled fermions are predicted to display exotic phenomena such as topological superfluidity and its associated Majorana fermions.  Here, we experimentally demonstrate a route to strongly-interacting single-component atomic Fermi gases by combining an $s$-wave Feshbach resonance (giving strong interactions) and spin-orbit coupling (creating an effective $p$-wave channel).  We identify the Feshbach resonance by its associated atomic loss feature and show that, in agreement with our single-channel scattering model, this feature is preserved and shifted as a function of the spin-orbit coupling parameters.
\end{abstract}


\maketitle

Feshbach resonances are characterized by singularities in two-body scattering, where a bound state becomes energetically degenerate with the two-body continuum.  For ultracold atoms these resonances are generally tuned with external magnetic fields that Zeeman-shift the energy difference between the bound (molecular) state and the two-atom continuum.  In two-component fermionic systems $s$-wave Feshbach resonances enabled the experimental realization of the crossover from Bardeen-Cooper-Schrieffer (BCS) pairing of fermions to the Bose-Einstein condensation (BEC) of bound molecules \cite{Regal2004,Bartenstein2004a,Zwierlein2004,Bourdel2004}.  The same crossover in spin-polarized Fermi gases involves a new kind of superfluid -- a topological superfluid -- with stable Majorana fermions at system edges and at internal defects \cite{Zhang2008, Ivanov2001}.  Owing to fermionic statistics, $s$-wave Feshbach resonances can only affect states of differing spin and therefore are absent in spin-polarized, single-component, Fermi gases.  $p$-wave Feshbach resonances do allow this coupling, but have a large associated loss which limits their utility \cite{Regal2003, Levinsen2008}.

In contrast it was recently shown that effective $p$-wave interactions arise from $s$-wave interactions when combined with spin-orbit coupling \cite{Zhang2008, Williams2012}.  A two-dimensional Fermi system with spin-orbit coupling is expected to exhibit chiral $p$-wave superfluidity with associated Majorana modes in vortices \cite{Zhang2008,Sato2009,Zhu2011,Gong2012}. Similarly, trapped one-dimensional spin-orbit coupled systems are predicted to support Majorana fermions at their boundary \cite{Jiang2011a,Liu2012, Wei2012}.  A large body of theoretical work focusses on the essential two-body physics of a Feshbach resonance in the presence of spin-orbit coupling and the resulting modification of the BEC-BCS crossover  \cite{Vyasanakere2011,Jiang2011b,Zhou2011,Ozawa2011,Yu2011,Li2012,He2012,Cui2012}.  To date, spin-orbit coupling has been experimentally realized in both Bose gases \cite{Lin2011b, Zhang2012, Qu2013}, and non-interacting Fermi gases \cite{Wang2012,Cheuk2012}.  In the latter case, radio-frequency spectroscopy near a Feshbach resonance has been reported \cite{Fu2013}.

In this paper, we experimentally study spin-orbit coupled $\K40$ atoms near a Feshbach resonance. Strong losses in atom number have been associated with Feshbach resonances in ultracold atoms since their initial observation \cite{Stenger99}.  We observe that the magnetic field value at which atom loss is maximum changes as a function of the strength and detuning of the Raman-dressing which generates the spin-orbit coupling.  We show that this change results from the energetically  shifted Raman-dressed continuum coming into resonance with a bound molecular state. This effect of the Feshbach resonance shifting with the spin-orbit coupling parameters has been predicted for the case of Rashba spin-orbit coupling \cite{Chaplik2006, Cappelluti2007, Vyasanakere2011, Gong2011, Yu2011, Shenoy2012}, while more recent theoretical work has considered the equal mixture of Rashba and Dresselhaus spin-orbit coupling considered here \cite{Shenoy2013, Melo2013}.

\begin{figure}[t]
\scalebox{1}{\includegraphics{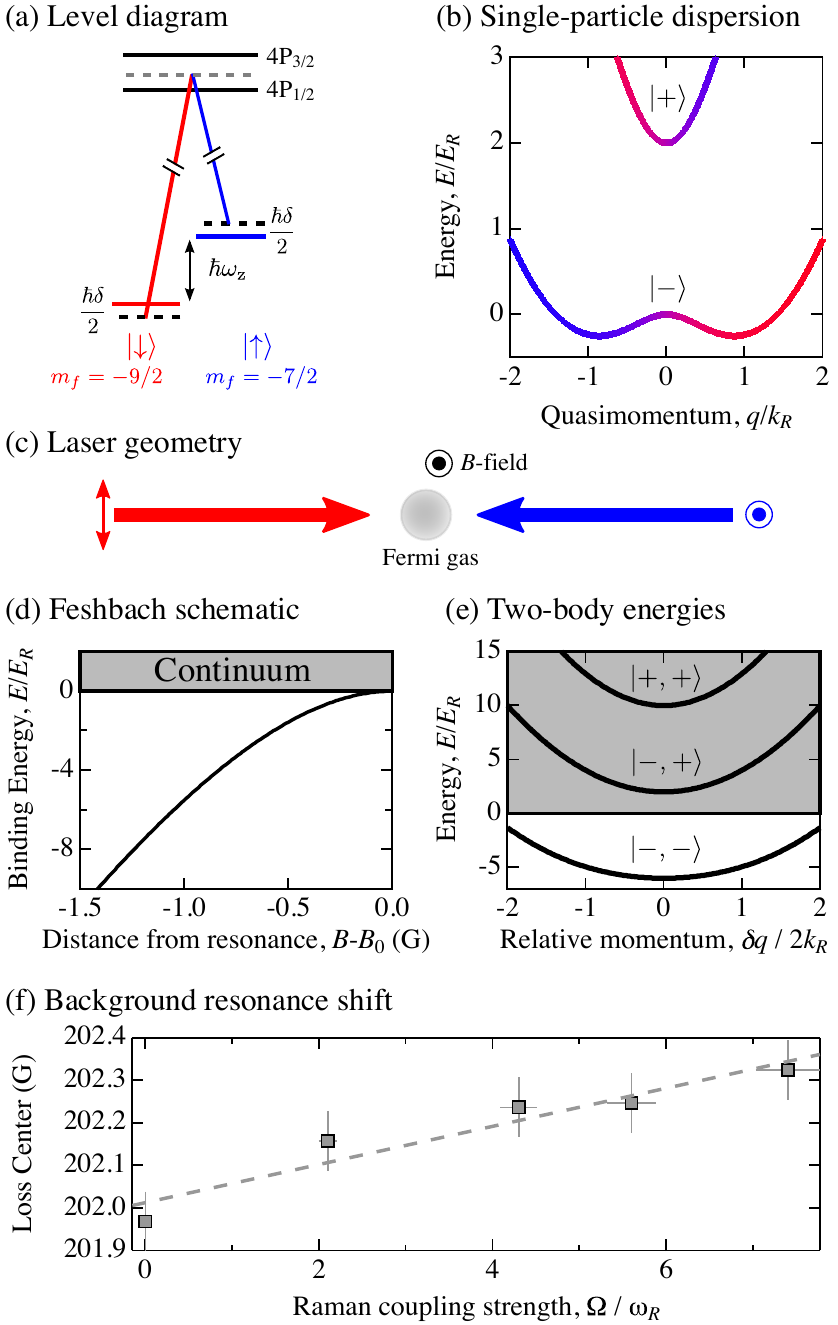}}
\caption{(a,c) Counter-propagating Raman beams couple two hyperfine spin states in $\K40$'s electronic ground state. (b) Energy bands for $\hbar\Omega = 2~\Er$, and $\delta=0$.  The color indicates the relative amplitude of the bare spin states in the dressed superposition. (d) Binding energy near the $\K40$ Feshbach resonance. (e)  Two-body energies showing the lowest band shift below the bare continuum for center-of-mass momentum $q_\mathrm{COM}=0$ and relative momentum $q_1-q_2 = 2\delta q$ for $\hbar\Omega = 8~\Er$ and $\delta=0$.  The upper and lower Raman-dressed bands are denoted using $\ket{+}$ and $\ket{-}$ respectively.  (f) Center of loss feature versus Raman coupling strength, $\Omega$ with a large two-photon detuning ($\hbar\delta=36~\Er$) illuminating an incoherent spin-mixture.}
\label{fig:setup}
\end{figure}

We generate spin-orbit coupling in a Fermi gas by dressing the $\ket{f=9/2,m_F=-7/2}\equiv\ket{\uparrow}$ and $\ket{f=9/2,m_F=-9/2}\equiv\ket{\downarrow}$ hyperfine spin states of $\K40$'s electronic ground state with a pair of counter-propagating Raman laser beams, Fig.~\ref{fig:setup}.  These spin states exhibit a broad Feshbach resonance at 202.1~G, with width $\simeq 7~\mathrm{G}$ \cite{Regal2004, Schneider2012}.  The Raman beams couple atoms in the $\ket{\uparrow,k=q+k_R}$ and $\ket{\downarrow,k=q-k_R}$ states~\cite{LinPRL2009} labeled by the quasimomentum $q$ that differ in linear momentum by $2\hbar\kr$, where $\hbar\kr = 2\pi\hbar\slash\lambda$ is the recoil momentum, and $\lambda=768.86$~nm is the Raman laser's wavelength.  The laser-dressed eigenstates $\ket{\pm,q}$ have a new single-particle dispersion relation consisting of two bands, shown in Fig.~\ref{fig:setup}(b), and are quasimomentum-dependent superpositions of $\ket{\uparrow}$ and $\ket{\downarrow}$.  The natural energy scale in the system is the recoil energy $\Er = \hbar\omega_R = \hbar^2\kr^2\slash 2m$, with $\omega_R = 2\pi\times 8.445~\mathrm{kHz}$.  The $\lambda=768.86~\mathrm{nm}$ wavelength of light providing the two-photon coupling is close to the zero-ac Stark shift magic wavelength such that the Raman beams contribute negligibly to the trapping potential.  The single-particle spin-orbit coupled system is completely characterized by the Raman coupling strength $\Omega$ and the two-photon detuning $\delta = \Delta\omega_L - \omega_\mathrm{Z}$, where $\Delta\omega_L$ is the angular frequency difference between the Raman beams and $\hbar\omega_\mathrm{Z}$ is the Zeeman splitting between the $\ket{\uparrow}$ and $\ket{\downarrow}$ states.  Note that the minimum of the lower Raman-dressed band is shifted downwards in energy as $\Omega$ increases; when $\Omega\gg4\Er$ this shift is $\approx \sqrt{(\Omega^2+\delta^2)}/2$.



Before studying the effect of Raman coupling on the Feshbach resonance, we carry out a control experiment to study the effect of the laser light independent of Raman dressing.
We create a spin-mixture of atoms in $\ket{\uparrow}$ and $\ket{\downarrow}$, illuminate the atoms with light that is far-detuned from Raman-resonance ($\hbar\delta=36 \Er$), and measure atom loss near the Feshbach resonance.  Surprisingly this off-resonant light shifts the loss feature to higher magnetic fields as shown in Fig.~\ref{fig:setup}(f).  This shift also exists for illumination by a single laser beam, confirming it is independent of the Raman-dressed physics of concern here.  We attribute this shift (effectively $45~\mathrm{mG}/\Er$) to a differential ac-Stark shift between the open channel atoms and molecular states (most likely the closed channel component) involved in the Feshbach resonance.  We find it exists over a range of wavelengths (while photoassocation-type effects associated with optical Feshbach resonances usually occur at a specific wavelength~\cite{Bauer2009a,Bauer2009b}, a recent experiment in $^{40}{\rm K}$ reported significant optical shifts of the $202.1\ {\rm G}$ resonance from bound-bound transitions~\cite{Fu2013a} over a range of wavelengths).  




At the most simple level, a Feshbach resonance occurs when two requirements are satisfied: (i) a pair of colliding atoms in the continuum must be degenerate with a molecular state, and (ii) the matrix element coupling the atomic and molecular states must be non-zero.  That the Raman laser coupling induces the Feshbach resonance and shifts its location relies on both of these facts: (i) the lower of the Raman-dressed bands is shifted downwards in energy, bringing potentially scattering atoms  into resonance with an otherwise bound molecule, and (ii) even within the same band, colliding atoms -- each of which is a momentum-dependent spin superposition -- couple to the molecular state by the usual $s$-wave coupling.  The finite momentum width of the Fermi gas ensures atoms exist in different superpositions of the bare spin states; while these fermions share the same ``dressed'' band index, they are microscopically non-identical and can interact with each other.

Sufficiently near a Feshbach resonance, the scattering takes on a universal form and only the last bound state approaching the continuum is relevant.  In this limit, the bound state's energy $E_B = \hbar^2/m a^2$ is simply related to the scattering length $a$ and the atomic mass $m$ \cite{Chin2010}.  Figure~\ref{fig:setup}(d) shows the energy of the last bound state computed for $\K40$ in the universal regime.  In comparison, Fig.~\ref{fig:setup}(e) shows the two-body Raman-dressed state energies, highlighting that Raman-dressed atoms in the lowest band lie below the initial continuum. We therefore anticipate a scattering resonance when the Raman-dressed continuum is degenerate with the initial molecular state.

The experiments begin with a spin-polarized gas of around $1.5\times10^5$ $\K40$ atoms in the $\ket{\downarrow}$ state at a temperature of $T\slash T_\mathrm{F} = 0.4$.  This degenerate Fermi gas is trapped in a crossed optical dipole trap with frequencies $\{\omega_x,\omega_y,\omega_z\}\slash 2\pi=\{39,42,134\}\,\mathrm{Hz}$.

A drawback of the current generation of experiments investigating spin-orbit coupling in alkali atoms is the spontaneous emission from off-resonantly excited atoms, leading to heating and loss.  Here this problem is mitigated by using a relatively fast (in the sense of not being limited by the trap frequency timescale) procedure to load fermions into the lowest Raman-dressed band.  We begin with a non-interacting spin-polarized Fermi gas in the $\ket{\downarrow}$ state at the desired magnetic field $B$, setting the distance from the Feshbach resonance.  We then ramp the Raman beams to their final power in $2~\mathrm{ms}$, slow compared to interband energy scales, but fast compared to the trap frequencies.  This loads the Fermi gas into the lowest Raman-dressed band with an average quasimomentum $q = \kr$, and thus a non-zero group velocity dependent on $\Omega$ and $\delta$.

Nearly all the experiments described in this paper share this procedure, after which the Raman-dressed fermions are held in the optical dipole trap for a variable time.  As the atoms move in the trap and change their quasimomentum, their internal state evolves adiabatically, for example upon reaching $q=0$ an atom is an equally weighted superposition of $\ket{\uparrow,q+\kr}$ and $\ket{\downarrow,q-\kr}$.  We consider two scenarios: (i) for $\Omega>4\Er$ we allow the Raman-dressed cloud to undergo a quarter-oscillation in the trap, at which point its center-of-mass momentum is zero; (ii) for $\Omega<4\Er$ the cloud initially sits in one-side of the double-well dispersion and has negligible initial group velocity.

\begin{figure}[t]
\scalebox{1}{\includegraphics{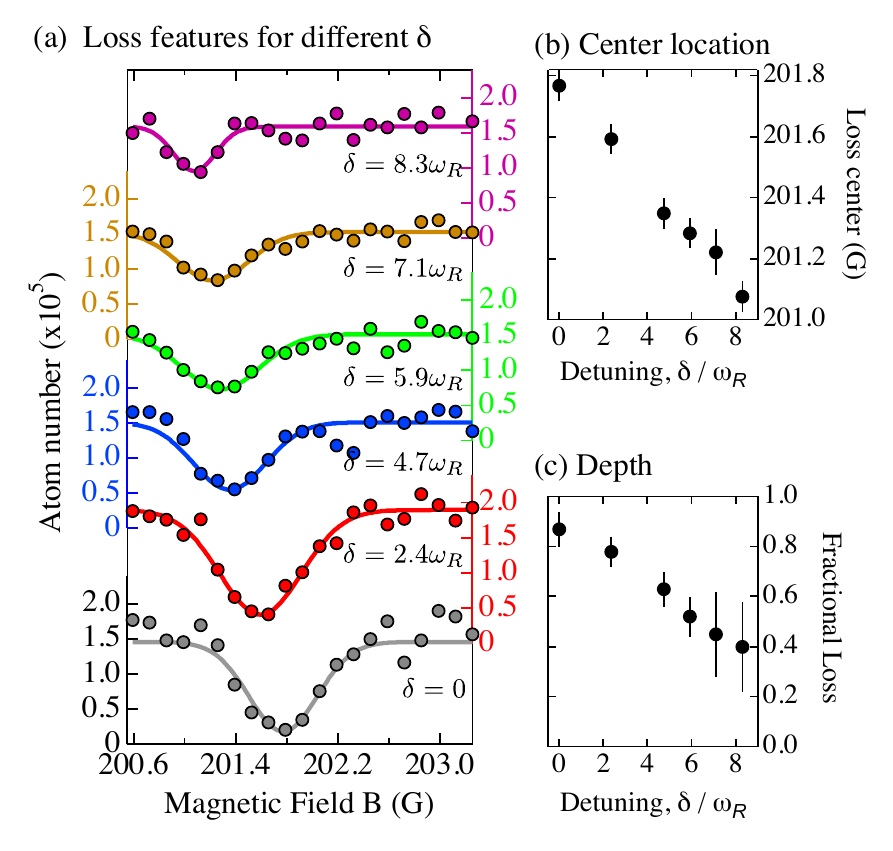}}
\caption{(a) Atom number as a function of magnetic field for $\hbar\Omega=4.7~\Er$ and detunings $\delta$: grey, $\hbar\delta=0$; red, $\hbar\delta=2.4~\Er$; blue, $\hbar\delta=4.7~\Er$; green, $\hbar\delta=5.9~\Er$; orange, $\hbar\delta=7.1~\Er$; magenta, $\hbar\delta=8.3~\Er$.  The solid curves are Gaussian fits to the data. (b) Center of loss feature as a function of $\delta$.  (c) Fraction of atoms lost as a function of $\delta$.  Uncertainty bars represent 1 standard deviation.}
\label{fig:detuning}
\end{figure}

We first investigate how the atomic loss feature of a spin-orbit coupled Fermi gas near a Feshbach resonance depends on parameters $\delta$ and $\Omega$.  Figure~\ref{fig:detuning}(a) shows the raw data from such experiments, in which the detuning $\delta$ is varied at fixed $\Omega$.  The different colored data sets represent different values of $\delta$.  The detuning is controlled at a given magnetic field (setting $\omega_\mathrm{Z}$) by changing the frequency difference $\Delta\omega_L$ between the Raman beams using acousto-optic modulators.

In this experiment, we keep the Raman coupling strength constant at $\hbar\Omega = 4.7~\Er$.  The initially moving Raman-dressed atoms are held in the dipole trap for $14~\mathrm{ms}$.  For $\hbar\Omega = 4.7~\Er$, $\delta=0$, this is the time the $\K40$ cloud takes to come to rest in the trap.  This hold time is longer than a quarter-period in the bare trap due to the modified dispersion relation of the dressed fermions.  After this hold time, the dipole trap is suddenly turned off ($<1~\mathrm{\mu s}$).  At the beginning of time-of-flight (TOF) the Raman-dressed atoms are mapped back to the bare spin state $\ket{\downarrow}$ by sweeping the detuning by an additional $\hbar\delta=20~\Er$ in $1~\mathrm{ms}$ and then turning off the Raman lasers in $1~\mathrm{ms}$.  We absorption-image the $\K40$ cloud after a total TOF of $6.3~\mathrm{ms}$ and count the remaining atoms.

We observe that with increasing $\delta$: (i) the absolute magnetic field value at which maximum loss occurs shifts to smaller values [Fig.~\ref{fig:detuning}(b), raw data, uncorrected for the off-resonant effect of Fig.~\ref{fig:setup}(f)]; (ii) the total atom loss in a fixed time decreases, Fig.~\ref{fig:detuning}(c).  The first observation can be explained by the downwards shifted Raman-dressed continuum coming into resonance with a bound molecular state.  The second observation results from the increasing spin-polarization of the dressed state (and hence reduced coupling to the molecular state) as $\delta$ increases.

\begin{figure}[t]
\scalebox{1}{\includegraphics{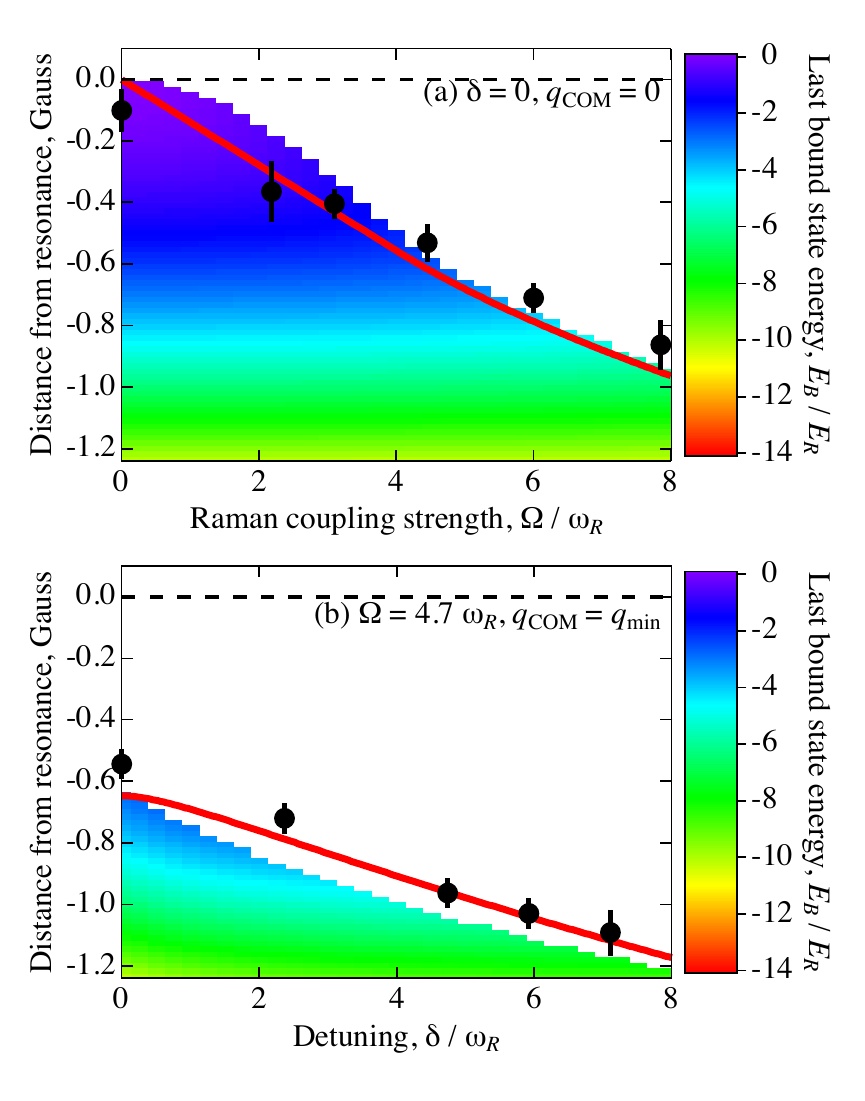}}
\caption{Predicted energy of the last bound molecular state as a function of: (a) Raman coupling strength $\Omega$ for $\delta=0$ and, (b) $\delta$ for $\hbar\Omega = 4.7~\Er$.  The color scale indicates the energy of the last bound state with center-of-momentum $\hbar q_\mathrm{COM}$, and the white region denotes where the bound state has entered the continuum and become a scattering state.  The red curve denotes the prediction of the simple model discussed in the text.  Experimental data, with the background shift of Fig.~\ref{fig:setup}(f) subtracted, is shown by the black circles.}
\label{fig:theory}
\end{figure}

To theoretically account for these results, we extend a single-channel scattering model to include Raman coupling between the two spin states.  We identify the shifted bound states from poles in the Green's function.  Figure~\ref{fig:theory} shows the calculated energy of the last molecular bound state (indicated by the color scale; the white regions signify that no molecule is present)  as a function of $\delta$ or $\Omega$ for a center of mass momentum $\hbar q_\mathrm{COM}$.  The red curve denotes the prediction of a simple model equating $E_B$ to the energy of a pair of atoms at the energy minimum of the lowest Raman-dressed band.

In a second measurement, we investigate loss near the Feshbach resonance as a function of the Raman coupling strength $\Omega$, with the detuning $\delta=0$.  As previously, we load the $\K40$ atoms into the lowest energy Raman-dressed band and hold the atoms in the optical dipole trap such that the Fermi gas only performs a single quarter-oscillation.  The hold times are between $10~\mathrm{ms}-15~\mathrm{ms}$ over the range of coupling strength we use.  After this hold time,  we determine the remaining number of atoms and map out the atomic loss signal as a function of magnetic field.

We compare the results of these experiments to theory in Fig.~\ref{fig:theory}, having now subtracted the intensity-dependent background shift of Fig.~\ref{fig:setup}(f).  The black symbols in Fig.~\ref{fig:theory}(a) plot the monotonically decreasing location of the loss feature as a function of Raman coupling $\Omega$, showing semi-quantative agreement with theory.  We determine the position of maximal atomic loss in the absence of spin-orbit coupling ($\Omega = 0$) using a spin-mixture rather than an initially spin-polarized Fermi gas.  Likewise Fig.~\ref{fig:theory}(b) depicts the loss feature moving to lower magnetic fields as a function of increasing detuning $\delta$, again in accord with theory.  In both of these cases, we load the Fermi gas into only the lowest Raman dressed state, where it is effectively a single-component system.  The continued effect of the Feshbach resonance in this polarized gas, is indicative of the effective $p$-wave interactions between these laser-dressed fermions -- just as laser-dressed bosons acquire effective $d$-wave interactions~\cite{Williams2012}.

\begin{figure}[t]
\scalebox{1}{\includegraphics{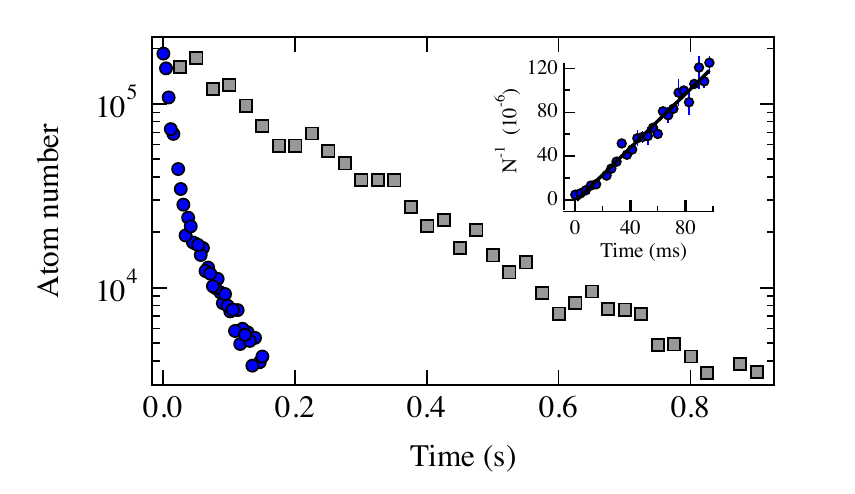}}
\caption{Atom number vs hold time in an optical dipole trap for different scenarios: (i) grey squares, $B=201.9~G$, $\hbar\Omega = 2.1~\Er$, $\hbar\delta = 33~\Er$; (ii) blue circles, $B=201.9~G$, $\hbar\Omega = 2.1~\Er$, $\hbar\delta = 0$.  Inset: the data of (ii) plotted as $1/N$.  The linear time dependence of $1\slash N$ is consistent with a two-body decay process.}
\label{fig:RamanCoupling}
\end{figure}

Finally we consider the rates associated with different loss mechanisms in the experiment.  Figure~\ref{fig:RamanCoupling} shows the number of atoms as a function of hold time in the optical dipole trap for two different scenarios.  The grey data displays the decay of a spin-polarized gas illuminated with off-resonant Raman beams ($\Omega = 2.1~\Er$, $\delta=33~\Er$).  In this case the atom loss is only due to the scattering of photons from the Raman lasers, giving a single-body decay with $1/e$ time of $180~\mathrm{ms}$ (the loss rate due to collisions with background atoms is negligible on the timescale of this experiment).  This comparatively slow rate implies that our $\approx14~\mathrm{ms}$ procedure for measuring loss is unaffected by spontaneous emission.

In contrast the blue data shows atom number for the on-resonant Raman beam case ($\Omega = 2.1~\Er$, $\delta=0$).  We load the atoms into the lowest dressed-band as described previously, at a magnetic field $B=201.9~\mathrm{G}$ (close to the field for maximum loss rate for this Raman coupling). The loss curve agrees well with that expected for a two-body loss process, as highlighted by the inset of Fig.~\ref{fig:RamanCoupling}(b) showing that the inverse of the atom number is linear in time.

Loss near a Feshbach resonance occurs in a two-step process whereby three-body collisions result in the formation of shallow dimers (which are not lost from the trap), followed by two-body collisions that allow the shallow dimers to decay to much more deeply bound molecules \cite{Zhang2011}.  The large kinetic energy released in this decay results in loss from the trap.  It is possible that the demapping (back to a single spin state) step in TOF (for which $\delta$ is increased by ramping $B$) dissociates the shallow dimers, allowing them to be imaged and counted.  This would make the loss rate resemble a two-body process, although two-body decay has previously been observed for fermions near a Feshbach resonance \cite{Dieckmann2002}.


In summary, we explored atomic loss in a spin-orbit coupled Fermi gas near a Feshbach resonance.  At sufficiently low temperatures this system in 1D (and possibly even 2D and 3D \cite{Seo2012}) is expected to allow topological superfluidity and support Majorana fermions.

\begin{acknowledgments}
We benefited greatly from conversations with C.~A.~R.~Sa~de~Melo, V. B. Shenoy, and E.~Tiesinga. This work was partially supported by the ARO with funding from DARPA's OLE program and the Atomtronics-MURI; and the NSF through the PFC at JQI. L.J.L. thanks NSERC and M.C.B. thanks the NIST-ARRA program.
\end{acknowledgments}


\end{document}